Research Papers

# Ni Foam Assisted Synthesis of High Quality Hexagonal Boron Nitride with Large Domain Size and Controllable Thickness

Hao Ying[1,2], Xiuting Li[1], Deshuai Li[3], Mingqiang Huang[4], Wen Wan[1], Qian Yao[1], Xiangping Chen[1], Zhiwei Wang[3], Yanqing Wu[4], Le Wang[2], and Shanshan Chen[2]*

[1] Department of Physics, Xiamen University, Xiamen, 361005, China
[2] Department of Physics, Renmin University of China, Beijing, 100872, China
[3] Beijing Institute of Nanoenergy and Nanosystems, Chinese Academy of Sciences, National Center for Nanoscience and Technology (NCNST), Beijing, 100083, China
[4] Wuhan National High Magnetic Field Center, Huazhong University of Science and Technology, Wuhan, 430074, China
* Corresponding author: Shanshan Chen, schen@ruc.edu.cn

## Abstract

The scalable synthesis of two-dimensional (2D) hexagonal boron nitride (h-BN) is of great interest for its numerous applications in novel electronic devices. Highly-crystalline h-BN films, with single-crystal sizes up to hundreds of microns, are demonstrated via a novel Ni foam assisted technique reported here for the first time. The nucleation density of h-BN domains can be significantly reduced due to the high boron solubility, as well as the large specific surface area of the Ni foam. The crystalline structure of the h-BN domains is found to be well aligned with, and therefore strongly dependent upon, the underlying Pt lattice orientation. Growth-time dependent experiments confirm the presence of a surface mediated self-limiting growth mechanism for monolayer h-BN on the Pt substrate. However, utilizing remote catalysis from the Ni foam, bilayer h-BN films can be synthesized breaking the self-limiting effect. This work provides further understanding of the mechanisms involved in the growth of h-BN and proposes a facile synthesis technique that may be applied to further applications in which control over the crystal alignment, and the numbers of layers is crucial.

# 1. Introduction

Hexagonal Boron Nitride (h-BN), a transparent two-dimensional (2D) atomic crystal, is the thinnest wide band gap insulator available, with high mechanical strength, chemical stability, and excellent thermal conductivity due to its strong in-plane covalent sp$^2$ bonds [1-5]. Thin h-BN layer has an atomically smooth surface free of dangling bonds and charge traps, which make it an ideal substrate for a variety of devices based on 2D materials, such as tunneling transistors [6-9] and Optical LED's[10]. Moreover, recent studies show graphene and h-BN hybrid structures exhibit fantastic properties including; band gap opening in graphene [11], Hofstadter's butterfly at high magnetic fields [12] , and a new breed quantum osscillations[13]. Besides, on its own h-BN has a wide range of applications such as ultraviolet-light emitter [1, 14], protective coatings [3] and heat dissipation [4, 5].

Similar to graphene, controllable synthesis of large area and high-quality h-BN with desirable thickness, is essential for both fundamental studies and technological applications. Novel methods such as physical vapor deposition (PVD) [15], atomic layer deposition (ALD) [16-19], ion beam sputtering deposition (IBSD) [20], and chemical vapor deposition (CVD) have been developed to synthesis the 2D h-BN films. So far, tremendous efforts have been made on large scale h-BN film synthesis utilizing various substrates such as Ni foil [21] , Cu foil [22-26], Cu-Ni alloys [27], Fe film [28], and Pt foil [29-31] via CVD. However, the CVD h-BN films reported are typically polycrystalline with small domains due to the high nucleation density at the early growth stage. These randomly oriented small domains lead to the high density of grain boundaries and dangling bonds, which dramatically degrade the mechanical and electronic properties of h-BN [14, 32]. Nucleation control for large, single crystalline h-BN domains has been achieved through different techniques, such as increasing the hydrogen gas flow [33], extending the annealing time of the substrate [26], electro-polishing the metal foil [24], folding the Cu foil into an enclosure [25], using Si-doped Fe catalyst [28] and adding Ni into Cu to make Cu-Ni alloy catalytic substrate [27]. Here we show a novel and effective technique to lower the nucleation density at the early growth stage of the h-BN and induce layer-by layer growth mode via Ni foam. The Pt foil was chosen as a catalytic substrate, given that extensive studies show it to be the best catalyst for the hydrogenation [34]. Kim *et. al.* further confirmed the superior catalytic property of Pt compare with Cu and Ru substrates for conversion growth of h-BN under the same reaction [35]. Moreover, by applying a bubbling-based method, rapid and nondestructive transfer of h-BN from Pt to arbitrary substrates and reuse of the Pt can be achieved to reduce environmental pollution as well as decrease the production cost of h-BN. Despite numerous advantages of the Pt foil, there has been sparse experimental work reporting the use of Pt [29-31] as an h-BN growth substrate in comparison of other metal substrates. To date, the largest reported h-BN domain size formed on polycrystalline Pt reaches 2 μm [24], which is one order smaller than that of the other metal substrates. Moreover, the growth dynamics of h-BN on Pt is, to date, poorly understood.

In this work, high-crystalline, large domain size h-BN was synthesized on Pt catalytic substrate via low pressure CVD (LPCVD) through the Ni foam assisted technique. The nucleation density of h-BN domains had been successfully lowered by over 2000 times on recyclable Pt. The crystalline structure of h-BN domain was found to be strongly related to the orientation of underlying Pt grains. Aligned h-BN domains were found not only on Pt (111), but also on Pt (100), Pt (110) and other high-index facets. The single crystalline h-BN films were only limited by the grain size of Pt substrate. The growth mode of the monolayer h-BN formed directly on Pt foil was confirmed to be surface-mediated and self-limiting. In addition, bilayer h-BN was formed on top of the first h-BN layer with the remote catalyst from Ni foam by increasing the heating temperature of ammonia borane (AB, $NH_3$-$BH_3$). Our technique provides a universal and facile route to synthesis high-quality h-BN films with large single crystals and controllable thickness.

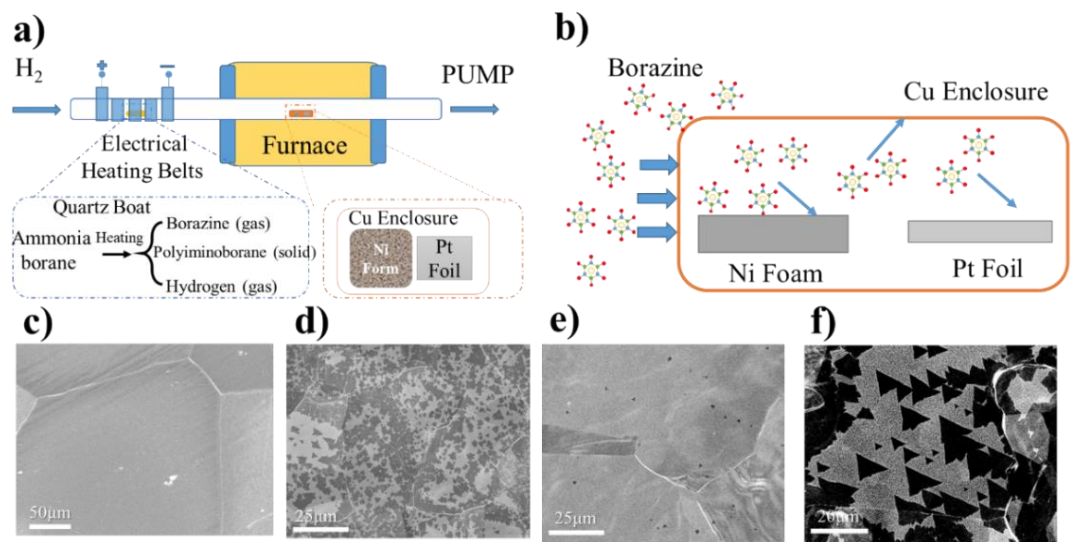

**Figure 1.** a) Schematic diagram of the LPCVD setup of the Ni foam assisted h-BN growth technique. b) Illustration depicts the mechanism that Ni foam absorbs most of the decomposed borazine, which effectively reduce the precursor feeding rate on the Pt foil in a highly confined space. c-e) SEM images of the h-BN films grown on the Pt foil only (c), Pt foil inside a Cu enclosure (d), Pt foil inserted into a Cu enclosure with a Ni foam (e). f) Typical aligned h-BN domains observed on Pt foil with the Ni foam assisted technique.

## 2. Results and discussion

Figure 1(a) illustrates schematically the LPCVD system applied for the growth of h-BN. AB was used as the precursor and placed in a quartz boat wrapped by a Cu foil to avoid the diffusion of the polyiminoborane nanoparticles residue decomposed from AB.[36] The substrate (a Pt foil inserted into a Cu enclosure together with the Ni foam on the upstream) was placed in the reaction zone. Typically, the growth of h-BN was performed at a temperature of 1050 °C with the AB precursor heated up to 75~100 °C (See Methods for more details). Due to the high specific surface area of the foam with 3D porous nanostructures (Figure S1, in the Supplementary Material) and the large solubility of boron atoms in Ni (~17at% at 1050 °C)[37, 38], the Ni foam has high capability to dehydrate and absorb the borazine decomposed from the AB precursor, and thus strongly reduces the precursor feeding rate onto the target substrate to control the nucleation density of h-BN. As depicted in Figure 1(b), the decomposed borazine was consumed mostly by the Ni foam, leaving small amounts of the borazine to be absorbed and decomposed by the Pt foil as well as the Cu enclosure. Figure 1(c-e) compare the morphology of the h-BN (under the same growth reaction) on the Pt foil only, on the Pt foil inside a Cu enclosure and on the Pt foil inside a Cu enclosure with a Ni foam with AB heated up to 75 °C for 30 min. It is clear that fully covered h-BN was observed on the Pt foil after 5 min growth, and by reducing the growth time down to 1 min, hundreds of nm of h-BN domain was seen on the Pt substrate (~ 17.8 nuclei/$\mu m^2$. Figure S1), while h-BN with lower nucleation density (~ 0.24 nuclei/$\mu m^2$) was formed on the Pt foil inside a Cu enclosure. This is consistent with the previous report that a decrease in the nucleation density was observed on the inner surface of the Cu enclosure.[25] A significant reduction of the nucleation density (~0.008 nuclei/$\mu m^2$) was further observed on the Pt foil inside a Cu enclosure with a Ni foam. The h-BN domain size up to ~22 μm was obtained on Pt foil with extended growth time, which is the largest single domain reported for h-BN growth on Pt foil[30] (Figure S1). A fully covered h-BN film could be achieved by increasing growth time.

The as-grown large domain h-BN films were then transferred onto the $SiO_2$/Si substrates, the Cu grid as well as the holey silicon nitride substrates (2.5 μm pores, PELCO, Ted Pella) for further characterization using a bubbling based transfer method[29]. Figure 2(a) shows the optical image of h-BN domains with a distinct triangular shape similar as observed on the Cu foil[22-26]. Well-aligned h-BN domains were found over a one-hundred-micron region. Those h-BN domains have a thickness of ~0.59 nm (inset of Figure 2(a)), consistent with a monolayer thickness on $SiO_2$ or $SiN_x$ (c-axis spacing of h-BN is ~0.32 nm)[22]. The uniform color contrast shown in Figure 2(b) indicates the thickness uniformity of the fully covered h-BN film. The AFM measurement shown in the inset of Figure 2(b) also confirms the monolayer thickness of the fully covered h-BN film. The surface roughness of the h-BN film was measured to be 0.107 nm, which is much lower than that of the $SiO_2$/Si substrate (0.143 nm), indicating of the smooth surface of the as-grown h-BN film (Figure S2). The measured surface roughness of h-BN film is comparable to the previously reported h-BN films[7, 25, 26]. The asymmetric Raman $E_{2g}$ mode at ~1371 $cm^{-1}$ obtained using a 488 nm laser (Figure 2(c)) confirms the growth of high-purity BN films with a hexagonal structure.[30] Moreover, the full width at half maximum (FWHM) of the Raman peaks is ~14.7 $cm^{-1}$, indicating the as-grown h-BN films are highly crystalline[24, 25, 30, 39]. Experimental work reported previously found that it was difficult to detect any Raman signals from the suspended 1L and 2L h-BN sheets[40]. However, when suspending our monolayer h-BN samples onto the holey silicon nitride substrates over the 2.5 μm holes, a strong Raman $E_{2g}$ mode at ~1372 $cm^{-1}$ was detected with a negligible third-order transverse optical mode from silicon substrate near 1450 $cm^{-1}$ (Figure 2(c))[41], further confirming the high quality of the as-grown h-BN films.

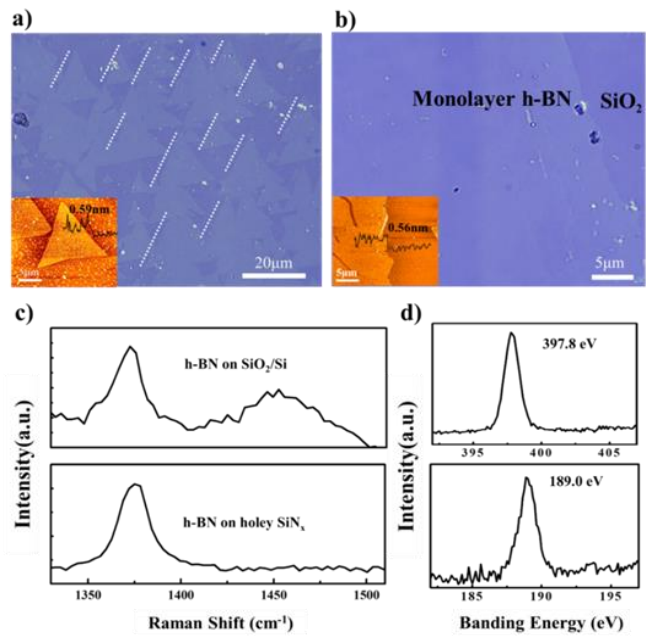

**Figure 2.** a) Optical image of the aligned h-BN domains transferred onto the 90 nm SiO$_2$/Si substrate. The inset shows the AFM image of the h-BN domain. The dash lines are the guide of eye to illustrate the well-aligned h-BN domains. b) Optical image of the monolayer h-BN film transferred onto the 90 nm SiO$_2$/Si substrate. The inset shows the corresponding AFM image of the monolayer h-BN. c) Raman spectrum of the supported and suspended monolayer h-BN film transferred onto the SiO$_2$/Si and holey silicon nitride substrate, respectively. d) XPS spectra of N1s and B1s of the h-BN film with binding energy peaks at 397.8 eV and 189.0 eV, respectively.

X-ray photoemission spectroscopy (XPS) was performed on the h-BN films that has been transferred on the SiO$_2$/Si substrate to determine the composition and stoichiometric ratio of B: N. The XPS spectra are shown in Figure 2(d), from which it can be determined that the B1s and N1s peaks are located at 189.0 eV and 397.8 eV, respectively. These values are consistent with those work previously reported for bulk boron nitride with a hexagonal phase[24-31]. The stoichiometric ratio of B and N atoms in the h-BN samples is calculated to be ~1.09 based on an XPS survey. Therefore, the as-synthesized films are dominantly composed of B-N bond and the local bonding of B and N atoms belongs to sp$^2$ hybridization. Moreover, no distinct peaks of the Pt or Ni impurities were detected on the samples through the XPS survey spectra (Figure S3).

Figure 3(a) shows the SEM morphology of h-BN domains over different Pt grains. The initial h-BN growth stage clearly shows different levels of coverage, as well as nuclei density of h-BN on different Pt grains where the Pt grain boundaries are clearly visible. Besides, h-BN domains are well-aligned within the Pt grain. In order to study the dependence of h-BN growth behavior on the crystalline orientation of the underlying Pt grains, electron backscatter diffraction mapping (EBSD) data were acquired over the sub-monolayer h-BN samples. As shown in Figure (3), three basic grain orientations (111), (110), (100) and other high index orientations were detected and mapped with EBSD on Pt foil. Almost fully covered h-BN film was observed over Pt (111) in Figure 3(a, d), while only a few h-BN nuclei appeared on Pt (100). Pt (110) and other high index surface show moderate h-BN coverages. The different h-BN coverages on the Pt lattice orientations could be attributed to the fact that the higher catalytic effect of the Pt (111) plane in comparison to other planes [31], which leads to the high efficient dehydrogenation of the borazine. The h-BN film, thus, preferentially forms on the Pt (111) surface, then the Pt (110) and the high index planes, and last the Pt (100) surface. Moreover, parallel-aligned h-BN domains were found not only on Pt (111), but also on Pt (100), Pt (110) and other high-index facets. Previously, aligned h-BN domains were reported selectively grown on Cu (111)[33] or on re-solidified Cu (110)[42] only by CVD. In this respect, Pt foil could be an ideal substrate for aligned h-BN growth. Figure 3(d) shows a lower magnification SEM image consisting of more Pt grains. The same colored dot line marks the aligned h-BN domains with the same oriented direction. Parallel-aligned h-BN domains with green dotted lines were observed in Pt grain 1 and grain 1' that have the same Pt orientation as indicated by the EBSD map shown in Figure 3(e). Well-aligned h-BN domains with yellow and purple dotted lines were observed also on Pt grain 2 and

2', Pt grain 3 and 3', respectively. Therefore, the well-aligned h-BN domains were not only observed within one Pt single crystalline grain but also in the adjacent and apart Pt grains with the same orientations. The EBSD map over the original Pt foil before growth was shown in Figure S4. It seems that the original Pt grains have more uniform orientation comparing with the one after growth (Figure 3(e)). Thus, the Pt grains are randomly evolving instead of having a fixed geometry and orientation during the h-BN growth under high temperature. This phenomenon has been observed previously on Cu foil for graphene growth by CVD[43]. Further studies will have to be performed for a better understanding of the evolution of Pt grain structures.

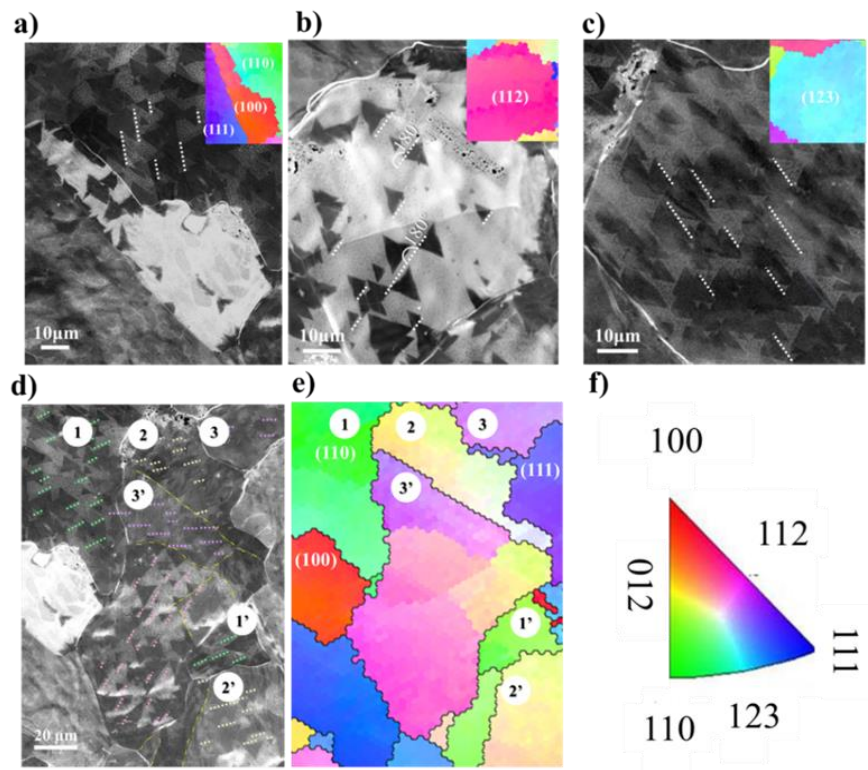

Figure 3. a-c) SEM images of the h-BN domains over various Pt grains. The insets show the corresponding EBSD maps. d-e) Lower magnification SEM image of the h-BN sample consists of more Pt grains and the corresponding EBSD map. f) The EBSD legend.

To confirm the crystalline structure of the well-aligned h-BN domains, we performed the transmission electron microscopy (TEM) on the transferred h-BN samples. Figure 4(a) shows a low magnification SEM image of the h-BN domains transferred onto a holey carbon grid. The marked triangles in the enlarged SEM images (Figures 4b, d-e) of region [1], [2] and [3] in Figure 4(a) clearly show the absence of those well-aligned h-BN domains. The selected area electron diffraction (SAED) patterns were obtained on various aligned h-BN grains labeled in region [1] (Figure 4(b)) and the corresponding patterns were shown in Figure 4(c). The SAED patterns obtained on the four chosen h-BN grains show the same characteristic 6-fold symmetric diffraction spot of h-BN materials, indicating the single-crystalline structure of the h-BN domains. The same set of SAED pattern confirmed that those well-aligned h-BN domains have the same crystalline structure. Furthermore, the h-BN domains in regions [2] and [3] share the same SAED pattern with those of region [1] (shown in the inset of Figure 4(d-e)), suggesting that the h-BN domains grown from different nucleation seeds are able to coalesce into a well-aligned film, and thus contribute to the high crystallinity h-BN films. These SAED patterns as well as high-resolution TEM (HR-TEM) image of a scroll-like edge (Figure 4(f)) reconfirm the monolayer nature of the as-grown h-BN film.

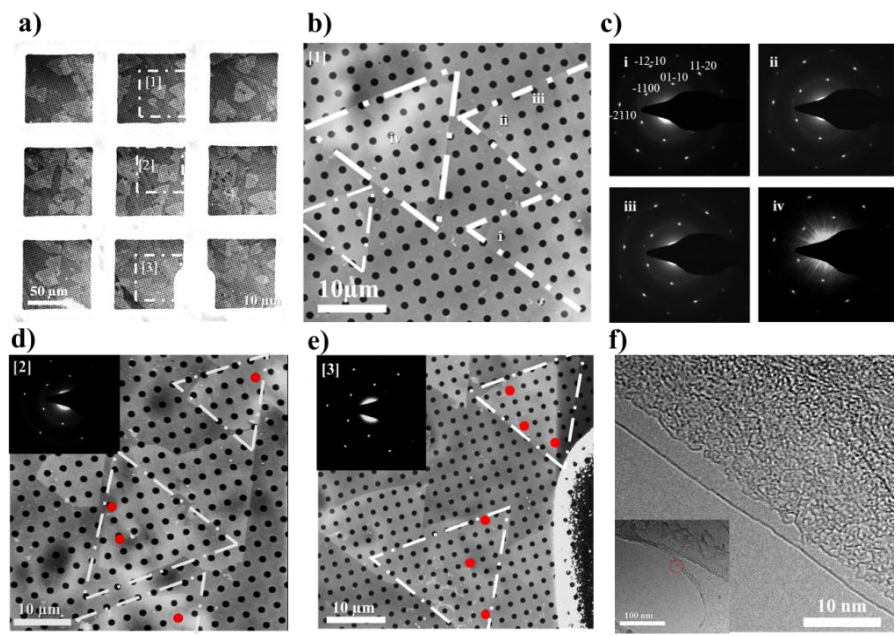

**Figure 4.** a) SEM image of the well-aligned h-BN domains transferred onto the holey carbon grid. b) SEM image of region [1] marked in (a). c) SAED patterns recorded on various spots labeled in (b). d-e) SEM images of region [2] and [3] marked in (a), respectively. The insets show the same set of SAED pattern recorded from the marked red spots, respectively. f) HR-TEM image of the h-BN domain edge on marked dashed circle region indicated in the inset.

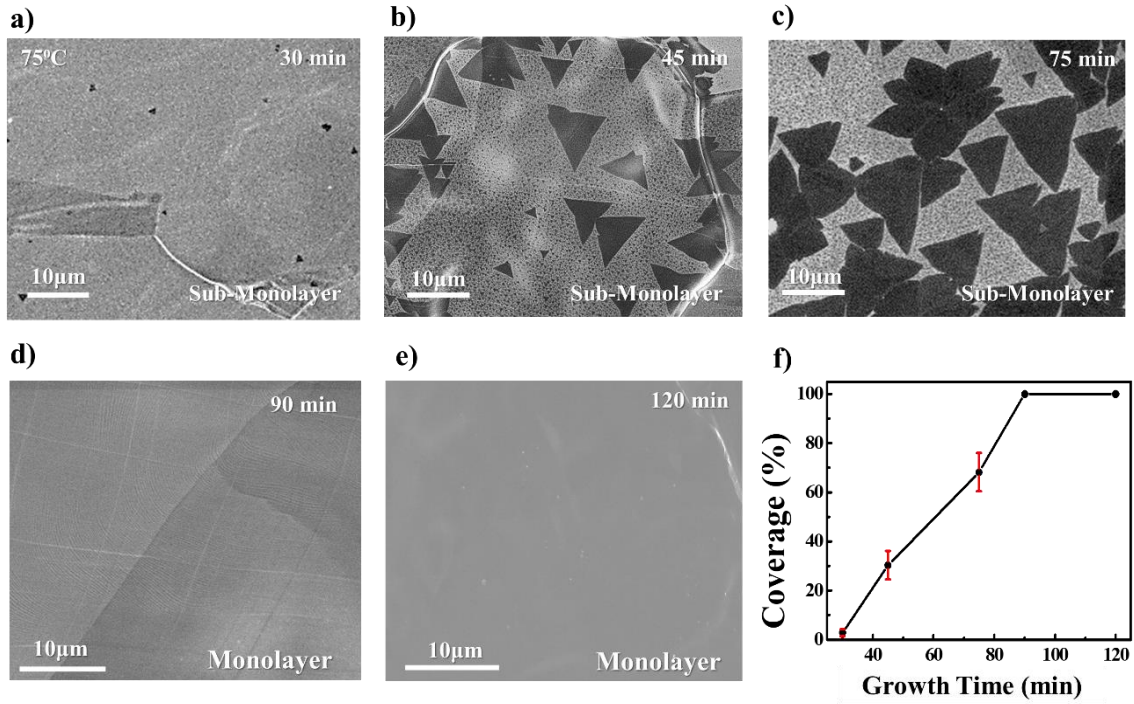

**Figure 5.** a-e) SEM images of the h-BN grown on Pt under $T_{AB}$ of 75 ℃ for 30 min (a), 45 min (b), 75 min (c), 90 min (d) and 120 min (e). f) Coverage of h-BN as a function of growth time.

Reported here for the first time is the initial nucleation and growth dynamics of h-BN grown on Pt substrates by LPCVD with increasing growth time. Figure 5(a) shows the sparse and small h-BN nuclei after 30 min growth, and those nuclei grow larger (Figure 5(b, c)) and then merge into a fully covered film after 90 min (Figure 5(d)). An extra 30 min was carried out with sufficient AB supply and no extra nuclei on top of the existing h-BN film were observed (Figure 5(e)). Further AFM measurement verifies the monolayer nature of the h-BN film grown for 90 min and 120 min, excluding the formation of uniform bilayer or multilayer h-BN after long growth time (Figure S5). The coverage of h-BN increased steadily with increasing growth time and then reached and stopped at the fully covered monolayer (Figure 5(f)). Therefore, the time dependent h-BN growth result confirms the surface mediated self-limiting growth mechanism of h-BN on Pt under LPCVD.

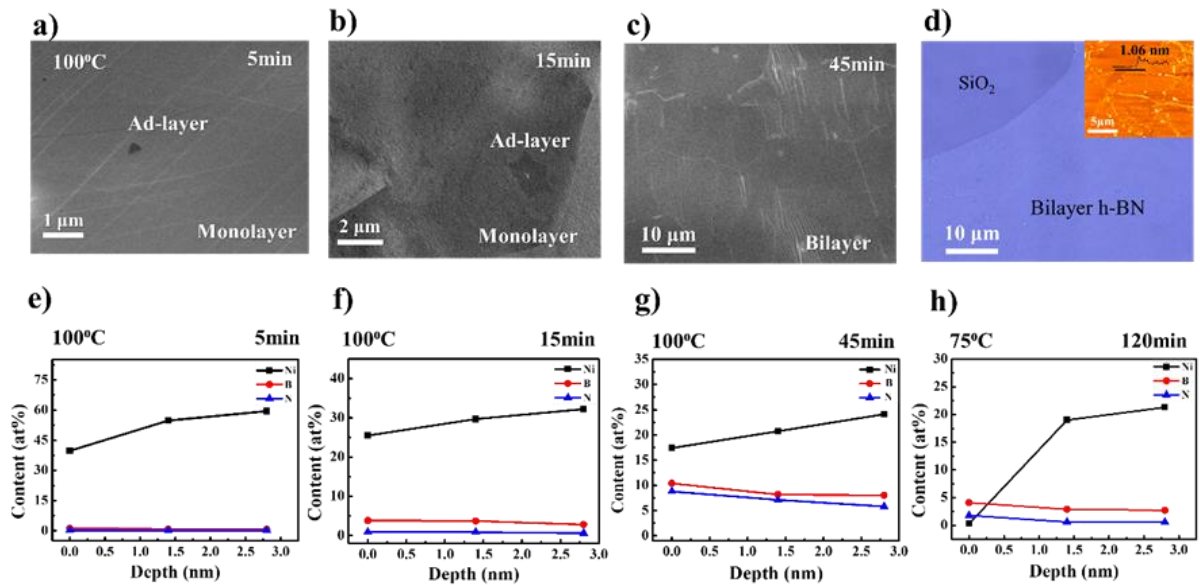

**Figure 6. a-c)** SEM images of the h-BN grown on Pt under $T_{AB}$ of 100 ℃ for 5 min (a), 15 min (b), and 45 min (c). d) Optical image of the bilayer h-BN film transferred onto the 90 nm SiO$_2$/Si substrate. The inset shows the corresponding AFM image. e-h) The AES depth analysis of the content of Ni, B, and N atoms on the Ni foams applied under $T_{AB}$ of 100 ℃ and the growth time of 5 min (e), $T_{AB}$ of 100 ℃ for 15 min (f), $T_{AB}$ of 100 ℃ for 45 min (g), and the $T_{AB}$ of 75 ℃ for 120 min (h).

In view of the high boron solubility as well as the high specific surface area of the Ni foam, we further explore the ability of Ni foam as an extra catalytic source to assist the growth of bilayer h-BN on Pt. The AB heating temperature ($T_{AB}$) was raised up to 100 $^0$C to speed up the growth of the first h-BN layer. The decomposition rate of AB increased dramatically and it led to the increase of the system pressure from 14 Pa to 16 Pa. Controlled experiments were carried out with and without the Ni foam to elucidate the role of Ni foam. The h-BN films grown without Ni foam show monolayer thickness after 5 min and 45 min growth (Figure S6), agreeing with the surface mediated self-limiting growth mechanism observed above. On the other hand, fully covered h-BN monolayer with some small h-BN ad-layers was obtained on Pt with Ni foam after 5 min as shown in Figure 6(a). It is quite interesting that growth of the first h-BN layer completes in 5 min both with and without the Ni foam with higher $T_{AB}$, which indicates that the absorption by the Ni foam does not play a crucial role under high $T_{AB}$. The ad-layer got enlarged (Figure 6(b)) and then merged into the bilayer h-BN (Figure 6(c)) with increasing growth time. The thickness of the as-grown h-BN films was measured to be 1.06 nm by AFM as shown in the inset of Figure6(d), confirming the formation of bilayer h-BN film. The Raman $E_{2g}$ mode at ~1370 cm$^{-1}$ and ~1371 cm$^{-1}$ were obtained on the transferred supported and suspended bilayer h-BN (Figure S7) which downshifts compared with the monolayer h-BN similar as reported previously[30, 40]. The HR-TEM image shows a folded edge of the bilayer h-BN (Figure S8), which exhibits two dark lines, reconfirms the bilayer nature of the as-grown h-BN film. In addition, one set of SAED pattern with the diffraction intensity of the outer {2110} peak much higher than the inner {1100} peak was recorded on the bilayer h-BN, indicating the high-symmetry stacking of the CVD-grown bilayer h-BN[44].

To verify the remote catalytic effect from the Ni foam, auger electron spectroscopy (AES) depth analysis was conducted on the Ni foams after h-BN growth. As shown in Figure 6 (e-g), the amount of B and N elements increases steadily with increasing growth time as higher coverage of h-BN are formed on the Ni foam surface and more B atoms are dissolved in Ni. The surface of Ni foam could be covered by h-BN when the Ni foam is saturated with B element, then the remote catalytic effect from Ni-foam stops. This could be seen from the AES depth profile of the monolayer h-BN grown under $T_{AB}$ of 75$^0$C for 120 min (Figure 6(h)), in which the Ni surface was fully covered by h-BN and nearly no Ni could be detected on the surface of the Ni foam to provide remote catalysis.

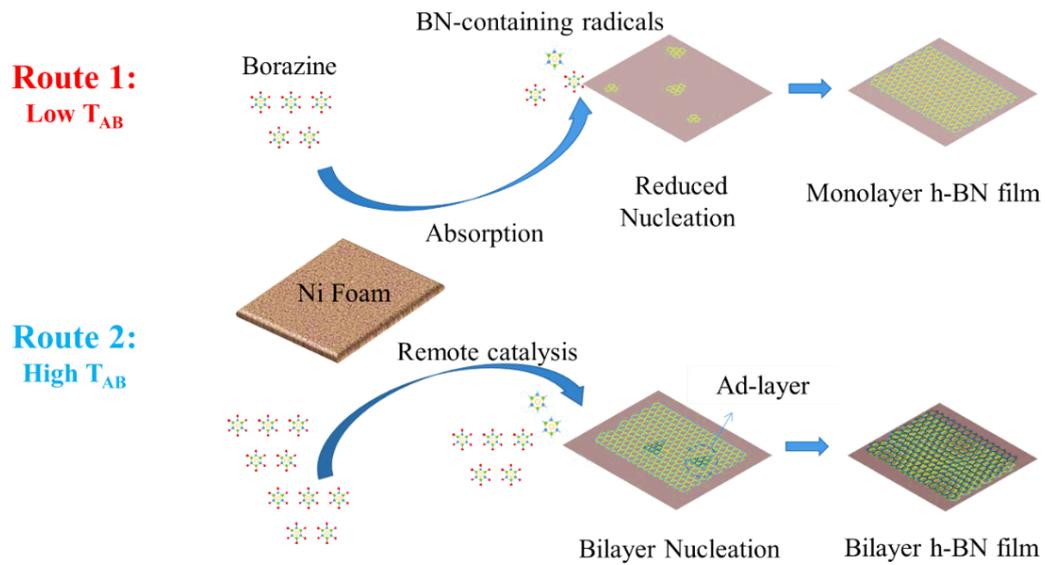

**Figure 7.** Schematic diagrams of temperature controllable switching growth of monolayer and bilayer h-BN film with the Ni foam assisted technique.

The mechanism of the Ni-foam assisted technique to grow the high-quality monolayer and bilayer h-BN is concluded in Figure 7. During low $T_{AB}$, the decomposed AB is mostly absorbed by the Ni foam which effectively lowered the precursor feeding rate onto the Pt foil in a highly confined space, and thus leads to a significantly reduced h-BN nucleation. The growth of fully covered h-BN on Pt takes longer than the saturated dissolution of B in the Ni foam. Therefore, only monolayer h-BN is formed on Pt surface through the self-limited surface adsorption process. With higher $T_{AB}$, the growth of monolayer h-BN on Pt completes in a few minutes and the Ni foam is far from saturated by sufficient B atoms. The Ni foam now acts as a remote catalyze substrate to assist the formation of second layer h-BN over the existing monolayer h-BN on Pt surface. Therefore, bilayer h-BN is able to form on the Pt substrate with the remote catalyst from Ni-foam, which breaks the self-limiting growth of monolayer h-BN on Pt during LPCVD.

We further fabricated graphene/h-BN and graphene/SiO$_2$ back gate field effect transistor (FET) devices to evaluate the quality of the CVD-grown h-BN (See Methods for more details). The electrical transport measurements were conducted in ambient condition. As shown in figure S9, the resistivity peak, corresponding to the overall charge neutrality point, appears at nearly zero voltage and ~16V for graphene/h-BN and graphene/SiO$_2$ devices, respectively. The FET device shows less p-type doping concentration and a significant enhancement of the carrier mobility (figure S9, 2198 and 1439 cm$^2$ V$^{-1}$s$^{-1}$,) with the h-BN support. These results imply that the high crystalline h-BN with much less charge impurities and carrier inhomogeneity than SiO$_2$ was produced.[7] The effectiveness of the CVD-grown h-BN as dielectric substrate for graphene devices is thus well proved.

## 3. Conclusions

In conclusion, we have developed a novel Ni foam assisted technique to synthesize high-crystalline monolayer and bilayer h-BN films. The nucleation density of h-BN domains has been successfully lowered by over 2000 times on recyclable Pt due to the high boron solubility as well as the high specific surface area of the Ni foam. Parallel-aligned h-BN domains were found on various Pt facets, demonstrating Pt as an ideal substrate for aligned h-BN growth. The lateral dimensions of the single-crystalline h-BN are up to hundreds of microns limited by the Pt grain size only. The growth of monolayer h-BN on Pt surface is confirmed to be self-limiting. With the remote catalysis from the Ni foam, bilayer h-BN film could be epitaxial grown on top of the first h-BN layer via a "layer by layer" mode to break the self-limit effect. These findings have provided a further understanding of the mechanisms involved in the growth of h-BN and propose a facile synthesis technique for other 2D materials as well as various heterostructure.

## 4. Methods

**4.1 LPCVD Growth of h-BN on Pt foils**

The Pt foil (99.99%, 100 μm thick) with a Ni foam on the upstream were inserted into a Cu enclosure, which then was loaded into a quartz tube furnace. Ammonia borane (97%, Sigma-Aldrich) precursor was put into a quartz boat wrapped by a Cu foil and heated to 75-100 ℃. The typical growth temperature is 1050 ℃. In the whole process, 10 sccm $H_2$ was used as the carrier gas under the pressure of 13 Pa. After growth, the sample was quickly cooled down to room temperature.

**4.2 Characterizations**

The as-grown h-BN samples were characterized by SEM (Zeiss Sigma), Raman spectroscopy (WItec alpha300, 488nm), EBSD (EDAX Inc) equiped with Sirion 200 SEM, TEM (FEI F20), XPS (PHI QUANTUM 2000), AFM (SPA 400) and AES (PHI 660).

**4.3 Devices fabrication and measurement**

Monolayer h-BN films were transferred onto the 300 nm $SiO_2$/Si substrate, followed by transferring of graphene flakes. For comparison, the graphene/h-BN and graphene/$SiO_2$ regions were chosen to fabricate the back gate FET devices using a standard electron beam lithography technique. The contact electrodes (5 nm Ti/30 nm Au) were deposited through electron beam evaporation. The electrical transport measurements were carried out in a home-made four-probe electrical measurement system. The gate voltage was applied between the back of the Si substrate and the drain electrode using a Keithley model 2400.

## Acknowledgements

We appreciate the support from the FANEDD through grant No. 201443, and from the NSFFPC through grant No. 2015J06016. S Chen also appreciates the supported by the Fundamental Research

**Supplementary Material:** Supplementary material is available online or from the author.